\documentclass{article}
\usepackage[center,footnotesize,hang]{subfigure}
\usepackage{graphicx}
\usepackage{amsmath}
\usepackage{bbm} 

\newcommand{\CenterEps}[2][1]{
\ensuremath{\vcenter{\hbox{\includegraphics[scale=#1]{#2.eps}}}}
}

\newcommand{\RaiseBrace}[1]{\raise3pt\hbox{$\displaystyle#1$}}

\def\ChargeC{\mathrm{C}}

\def\SO{\text{SO}}

\def\<{\left\langle}
\def\>{\right\rangle}

\def\be{\begin{equation}}
\def\ee{\end{equation}}
\def\beq{\begin{equation}}
\def\eeq{\end{equation}}
\def\bea{\begin{eqnarray}}
\def\eea{\end{eqnarray}}

\begin{document}
 
\baselineskip 18pt

\begin{center}
{\Large {\bf  
Neutrino Mass Models and Leptogenesis
}}
 \vskip5mm S.~F.~King$^{1,}$\footnote{Email address:
   sfk@hep.phys.soton.ac.uk}\\
$^{1}$School of Physics and Astronomy,
University of Southampton,\\
Southampton, SO17 1BJ, United Kingdom, \\
E-mail: sfk@hep.phys.soton.ac.uk
\bigskip

\begin{abstract} 
In this talk
we show how a natural neutrino mass hierarchy can follow from
the type I see-saw mechanism, and a natural neutrino mass 
degeneracy from the type II see-saw mechanism, where the bi-large
mixing angles can arise from either the neutrino or charged lepton
sector. We summarize the phenomenological implications
of such natural models, and discuss the model building applications
of the approach, focussing on the $SU(3)\times SO(10)$ model. 
We also show that
in such type II models the leptogenesis asymmetry parameter becomes
proportional to the neutrino mass scale, in sharp contrast to the
type I case, which leads to an upper bound on the neutrino mass scale,
allowing lighter right-handed neutrinos and hence making leptogenesis
more consistent with the gravitino constraints in supersymmetric
models.

\end{abstract}
\end{center}
\vspace{5.0in}
\noindent Based on invited talks presented at the 
10th International Symposium on Particles, Strings and Cosmology
(Pascos04), Northeastern University, Boston, August 16-22, 2004
and Nobel Symposium 129 on Neutrino Physics,
Haga Slott, Enkoping, Sweden, August 19-24, 2004.

\newpage

\section{Introduction}

The discovery of neutrino mass and mixing at the end of the last
century implies that the Standard Model is incomplete and needs
to be extended, but how \cite{King:2003jb}? 
In attempting to answer this question, 
it is useful to being by classifying models in terms of the mechanisms
responsible for small neutrino mass, and large lepton mixing,
as a first step towards finding the Next Standard Model.
Amongst the most elegant mechanisms for small neutrino
mass is the see-saw mechanism \cite{seesaw}.
However the see-saw mechanism by itself does not provide
an explanation for bi-large lepton mixing for 
either hierarchical or denegerate neutrinos.

In this talk we discuss model independent approaches 
to accounting for bi-large mixing in a natural way,
based on the see-saw mechanism, which are valid for
both hierarchical or denegenerate neutrino mass spectra.
For the case of hierarchical neutrino masses arising from the
type I see-saw mechanism, it is shown how the neutrino mass
hierarchy and bi-large mixing angles could originate
from the sequential dominance of right-handed neutrinos 
\cite{King:1998jw}. It is then shown how
to obtain partially degenerate neutrinos
in a natural way by including a type II contribution
proportional to the unit mass matrix, 
with the neutrino mass splittings and
mixing angles controlled by type I contributions
and sequential dominance \cite{Antusch:2004xd}.
The bi-large mixing angles could originate either from
the neutrino or the charged lepton sector \cite{Antusch:2004re}.
For a review see \cite{Antusch:2004gf}. 
We summarize the phenomenological implications
of such natural models, and discuss the model building applications
of the approach, focussing on the $SU(3)\times SO(10)$ model. 
We also discuss leptogenesis 
in such type II models.
The leptogenesis asymmetry parameter becomes
proportional to the neutrino mass scale, in sharp contrast to the
type I case, which leads to an upper bound on the neutrino mass scale,
allowing lighter right-handed neutrinos and hence making leptogenesis
more consistent with the gravitino constraints in supersymmetric
models \cite{Antusch:2004xy}.

\section{See-Saw Mechanism}
\begin{figure}
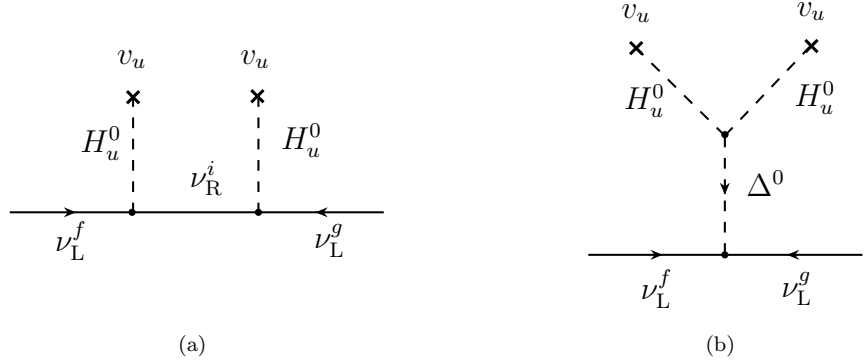

\begin{center}
  \subfigure[\label{fig:TypeIDiagrams}]{$ 
 \CenterEps[1]{typeIseesaw}\vphantom{\CenterEps[1]{typeIIseesaw}}
  $}  \hfil
  \subfigure[\label{fig:TypeIIDiagrams}]{$ 
  \CenterEps[1]{typeIIseesaw}
  $}
 \caption{Diagram (a) shows the contribution from the exchange of a heavy
right-handed
neutrino as in the type I see-saw mechanism. Diagram (b) illustrates
the
 contribution from an induced vev of
the triplet $\Delta$. At low energy, they can be viewed as
contributions to
the effective neutrino mass operator from
integrating out the heavy fields $\nu_{\mathrm{R}}^i$ and $\Delta^0$,
respectively.
 }
\end{center}
\end{figure}

The most commonly discussed version of the see-saw mechanism
is sometimes called the type I see-saw mechanism \cite{seesaw}. 
The type I see-saw mechanism is illustrated diagramatically in 
Fig.~\ref{fig:TypeIDiagrams}.

In models with a left-right 
symmetric particle content like minimal left-right 
symmetric models, Pati-Salam models or grand unified theories (GUTs) 
based on $\SO (10)$, the type I see-saw mechanism is often  
generalized to a type II see-saw (see e.g.~\cite{Lazarides:1980nt}), 
where an additional direct 
mass term $m_{\mathrm{LL}}^{\mathrm{II}}$ for the light neutrinos is present. 

With such an additional direct mass term, 
the general neutrino mass matrix is given by
\begin{eqnarray}
\,
\left( \begin{array}{cc} \overline{\nu_{\mathrm{L}}}  &
  \overline{\nu^{\ChargeC }_\mathrm{R}}   \end{array}   \right) 
\left( \begin{array}{cc} 
m^{\mathrm{II}}_{\mathrm{LL}} \vphantom{\nu_{\mathrm{L}}^{\ChargeC }}&
m^\nu_{\mathrm{LR}}\\[1mm]
 m^{\nu T}_{\mathrm{LR}} \vphantom{\nu_{\mathrm{L}}^{\ChargeC }}&
 M_{\mathrm{RR}}
    \end{array}   \right)
  \,
 \left( \begin{array}{c} 
\nu_{\mathrm{L}}^{\ChargeC }  \\[1mm]
  {\nu_\mathrm{R}}   \end{array}   \right) .
\end{eqnarray}
Under the assumption that the mass eigenvalues $M_{\mathrm{R}i}$ of
$M_{\mathrm{RR}}$ are very large compared to the components of  $m^{\mathrm{II}}_{\mathrm{LL}}$
and $m_{\mathrm{LR}}$, the mass matrix can approximately be 
diagonalized yielding effective Majorana masses  
\begin{eqnarray}\label{eq:TypIMassMatrix}
m^\nu_{\mathrm{LL}} \approx 
m^{\mathrm{II}}_{\mathrm{LL}} + m^{\mathrm{I}}_{\mathrm{LL}}
\end{eqnarray} 
with 
\begin{eqnarray}
m^{\mathrm{I}}_{\mathrm{LL}} \approx - m^\nu_{\mathrm{LR}}
\,M^{-1}_{\mathrm{RR}}\,m^{\nu T}_{\mathrm{LR}}
\end{eqnarray}
for the light neutrinos. 
The direct mass term  
$m^{\mathrm{II}}_{\mathrm{LL}}$ can also provide a naturally small contribution to the 
light neutrino masses if it stems e.g.~from a see-saw suppressed induced vev. 
The type II contribution may be induced
via the exchange of heavy Higgs
triplets of SU(2)$_\mathrm{L}$ as illustrated diagramatically in 
Fig.~\ref{fig:TypeIIDiagrams}.

\section{A Natural Neutrino Mass Hierarchy}

In this section we discuss an elegant and natural way of
accounting for a neutrino mass hierarchy and two large mixing angles
in the type I see-saw mechanism. 
The starting point is to assume that one of
the right-handed neutrinos contributes dominantly to the
see-saw mechanism and determines the atmospheric neutrino
mass and mixing. A second right-handed neutrino
contributes sub-dominantly and determines the solar neutrino
mass and mixing. The third right-handed neutrino
is effectively decoupled from the see-saw mechanism.

The above Sequential Dominance mechanism \cite{King:1998jw}
is most simply described
assuming three right-handed neutrinos
in the basis where the right-handed neutrino mass matrix is diagonal
although it can also be developed in other bases.
In this basis we write the input
see-saw matrices as
\begin{equation}
M_\mathrm{RR}=
\left( \begin{array}{ccc}
X & 0 & 0    \\
0 & Y & 0 \\
0 & 0 & Z
\end{array}
\right), \ \ \ \ 
m^\nu_\mathrm{LR}=
\left( \begin{array}{ccc}
a & d & p    \\
b & e & q\\
c & f & r
\end{array}
\right)
\label{seq1}
\end{equation}
Each right-handed neutrino in the basis of Eq.\ref{seq1}
couples to a particular column of $m^\nu_\mathrm{LR}$.
There is no mass ordering of $X,Y,Z$ implied in Eq.\ref{seq1}.
The dominant right-handed neutrino may be taken to be the one
with mass $Y$ without loss of generality.
Sequential dominance occurs when the
right-handed neutrinos dominate sequentially \cite{King:1998jw}:
\beq \label{srhnd}
\frac{|e^2|,|f^2|,|ef|}{Y}\gg
\frac{|xy|}{X} \gg
\frac{|x'y'|}{Z}\; ,
\eeq
where $x,y\in a,b,c$ and $x',y'\in p,q,r$.
This leads to a full neutrino mass hierarchy
$m_3^2\gg m_2^2\gg m_1^2$. Ignoring phases,
in the case that $d=0$, corresponding to a Yukawa 11 texture zero
in Eq.\ref{seq1}, we have:
\beq
m_1  \sim  O(\frac{x'y'}{Z}),  \ \ \ 
m_2  \approx   \frac{|a|^2}{Xs_{12}^2},  \ \ \ 
m_3  \approx  \frac{|e|^2+|f|^2}{Y}
\label{masses}
\eeq 
where $s_{12}=\sin \theta_{12}$ is given below.  Note that 
each neutrino mass is generated by a separate right-handed
neutrino, and the sequential dominance condition naturally results in
a neutrino mass hierarchy $m_1\ll m_2\ll m_3$.  The neutrino mixing
angles are given to leading order in $m_2/m_3$ by \cite{King:1998jw}:
\beq
\tan \theta_{23}  \approx \frac{e}{f}, \ \ 
\tan \theta_{12}  \approx  \frac{a}{c_{23}b-s_{23}c}, \ \  
\theta_{13}  \approx 
s_{12}c_{12} \frac{(s_{23}b+c_{23}c)}{(c_{23}b-s_{23}c)}\frac{m_2}{m_3}
\label{angles}
\eeq
Physically these results show that in sequential dominance 
the atmospheric neutrino mass $m_3$ and mixing $\theta_{23}$
is determined by the
couplings of the dominant right-handed neutrino of mass $Y$.
The angle $\theta_{13}$ is generically of order 
$\theta_{13}\sim O(m_2/m_3) \sim 0.2$.
However the coefficient in Eq.\ref{angles} can be
arbitrarily small, since to leading order as 
$b\rightarrow -c$, $\theta_{13}\rightarrow 0$, but $\theta_{12}$
remains large.
The solar neutrino mass $m_2$ and mixing $\theta_{12}$
is determined by the couplings of the
sub-dominant right-handed neutrino of mass $X$.
From Eq.\ref{angles}, the solar angle only depends
on the sub-dominant couplings and
the simple requirement for large solar angle is $a\sim b-c$.
The third right-handed neutrino of mass $Z$
is effectively decoupled from the see-saw mechanism
and leads to the vanishingly small mass $m_1\approx 0$.

\section{A Natural Neutrino Mass Degeneracy}
We now show that it is possible to obtain 
a (partially) degenerate neutrino mass spectrum by essentially adding
a type II direct neutrino mass 
contribution proportional to the unit matrix:
Thus we shall consider a type II extension 
\cite{Antusch:2004xd}, where the mass matrix of the light neutrinos 
has the form:    
\begin{eqnarray}
m^\nu_{\mathrm{LL}} \approx m^{\mathrm{II}}\, 
\left(\begin{array}{ccc}
1&0&0\\
0&1&0\\
0&0&1
\end{array}\right)
 +
m_{\mathrm{LL}}^{\mathrm{I}}
\end{eqnarray}  
Assuming here that
 the type I mass matrix $m_{\mathrm{LL}}^{\mathrm{I}}$ is
 real, the full neutrino mass matrix is diagonalised by the
same matrix that diagonalises the type I matrix:
  \begin{eqnarray}\label{eq:TypIISeeSawFormulaUnitMatrix}
 (m^\nu_{\mathrm{LL}})_{\mathrm{diag}} \;= \; 
 m^{\mathrm{II}}\, V V^T +  V  m_{\mathrm{LL}}^{\mathrm{I}} V^T
\;=\; m^{\mathrm{II}}\, \mathbbm{1} + 
diag(m_1^{\mathrm{I}},m_2^{\mathrm{I}},m_3^{\mathrm{I}}).  
 \end{eqnarray}
In this case the
neutrino mass scale is controlled by the type II mass scale 
$m^{\mathrm{II}}$,
while the neutrino mass splittings are determined by the type I mass
eigenvalues:
\beq 
m_1  \approx  |m^{\mathrm{II}}- m_2^{\mathrm{I}} |, \ \ \ 
m_2  \approx | m^{\mathrm{II}} - m_2^{\mathrm{I}}|, \ \ \
m_3  \approx | m^{\mathrm{II}} - m_3^{\mathrm{I}}|.
\eeq
Sequential dominance in the type I sector naturally predicts
$m_1^{\mathrm{I}}\ll m_2^{\mathrm{I}}\ll m_3^{\mathrm{I}}$.
Hence the very small mass splittings required for degenerate
neutrinos can be achieved naturally by sequential dominance.
The predictions for the mixings are determined from the type I 
mass matrix (the type II unit matrix is irrelevant).
In particular the atmospheric and solar angles are given
by the sequential dominance estimates in
Eq.\ref{angles} and are independent
of the type II neutrino mass scale $m^{\mathrm{II}}$.
However the angle $\theta_{13}$ in Eq.\ref{angles} is now of order 
$\theta_{13}\sim { O}(m_2^{\mathrm{I}}/m_3^{\mathrm{I}})$.
This is now much smaller than the type I result
$\theta_{13}\sim { O}(m_2/m_3)$
since the neutrino mass splittings, controlled by the type I masses,
are very small for partially degenerate neutrinos.

\section{Mixing Angles From the Charged Leptons?}\label{sec:SmallNuMixing}

In this section we show how bi-large mixing could originate from
the charged lepton sector using a 
generalization of sequential right-handed neutrino dominance to all
right-handed leptons \cite{Antusch:2004re}. 
We write the mass matrices for the charged leptons  
$m_\mathrm{E}$ as 
\begin{eqnarray}
m_\mathrm{E} = \left(
\begin{array}{ccc}
p' & d' & a' \\
q' & e' & b' \\
r' & f' & c'
\end{array}
\right)\!.
\end{eqnarray}
In our notation, 
each right-handed charged lepton couples to a column in
$m_\mathrm{E}$.
For the charged leptons, the sequential dominance conditions are
\cite{Antusch:2004re}:
\begin{eqnarray}\label{eq:SeqDominanceCL}
|a'|,|b'|,|c'| \gg |d'|,|e'|,|f'| \gg |p'|,|q'|,|r'| \;.
\end{eqnarray}
They imply the desired hierarchy for the charged lepton masses $m_\tau
\gg m_\mu \gg m_e$ and small right-handed mixing of
$U_{e_\mathrm{R}}$. 
We assume zero mixing from the neutrino sector.
A natural possibility 
for obtaining a small $\theta_{13}$ is \cite{Antusch:2004re}
\begin{eqnarray}\label{eq:CondForSmallT13}
|d'|,|e'| \ll |f'| \; .
\end{eqnarray} 
In leading order in $|d'|/|f'|$ and $|e'|/|f'|$, 
for the mixing angles $\theta_{12},\theta_{23}$ and $\theta_{13}$, we obtain   
(again ignoring phases here)
\beq
\tan \theta_{12} \approx  \frac{a'}{ b'}, \ \ 
\tan \theta_{23}\approx   \frac{  s_{12}\, a' +  c_{12}
\,b' }{c' }, \ \ 
\tan \theta_{13} \approx \frac{s_{12}\,e' \, - c_{12}d' \, }{f'\,}  
\eeq

$\theta_{13}$ only depends on $d'/f'$ and $e'/f'$ from 
the Yukawa couplings to the sub-dominant right-handed muon and on $\theta_{12}$.  
We find that in the limit $|d'|,|e'| \ll |f'|$, the two large mixing angles 
$\theta_{12}$ and $\theta_{23}$ approximately depend only on 
$a'/c'$ and $b'/c'$ from 
the right-handed tau Yukawa couplings.  
Both mixing angles are large if $a',b'$ and $c'$ are of the same order.

\section{Phenomenological Implications }
We now summarize the
phenomenological consequences of type I see-saw models with 
a natural mass hierarchy and their type II extensions with a natural
mass degeneracy, for the low energy 
neutrino parameters and high-energy mechanisms as leptogenesis and 
minimal lepton flavour violation (LFV).  
In order to compare the predictions of such 
natural see-saw models (based on sequential dominance) 
with the experimental data obtained at low energy, the 
renormalization group (RG) running of the effective neutrino
mass matrix has to be taken into account.   

\subsection{Renormalization Group Corrections}\label{sec:RGRunning}
For type I models with sequential dominance, the running of the mixing angles 
is generically small \cite{King:2000hk} 
since the mass scheme is strongly hierarchical.  
When the neutrino mass scale is lifted, e.g.\ via a type II upgrade, 
a careful treatment of
the RG running of the neutrino parameters, 
including the energy ranges between and above the see-saw
scale \cite{King:2000hk}, is required. 
Dependent on $\tan \beta$ in the MSSM, on the size of the 
neutrino Yukawa couplings and on the neutrino mass scale, the RG effects can  
be sizable or cause only small corrections.

\subsection{Dirac and Majorana CP Phases and Neutrinoless Double Beta Decay} 
At present, the CP phases in the lepton sector are unconstrained by
experiment.  In type I see-saw models based on sequential dominance,
there is no restriction on them from a theoretical point of view.  The
type-II-extension however predicts that all observable CP phases,
i.e.~the Dirac CP phase $\delta$ relevant for neutrino oscillations
and the Majorana CP phases $\beta_2$ and $\beta_3$, become small as
the neutrino mass scale increases.

The key process for measuring the neutrino mass scale could be
neutrinoless double beta decay. The decay rates depend on an effective
Majorana mass defined by $\< m_\nu \>= \left| \sum_i
(U_\mathrm{MNS})_{1i}^2 \, m_i \right|$.  Future experiments which are
under consideration at present might increase the sensitivity to $\<
m_\nu \>$ by more than an order of magnitude. 
For type I models with
sequential dominance, which have a hierarchical mass scheme, $\< m_\nu
\>$ can be very small, below the accessible sensitivity.

For models where the neutrino mass scale is lifted via a type II extension 
\cite{Antusch:2004xd}, there
is a close relation between the neutrino mass scale, i.e.\ the mass of
the lightest neutrino and $\< m_\nu \>$. Since the CP phases are
small, there can be no significant cancellations in $\< m_\nu \>$.  This
implies that the effective mass for neutrinoless double beta decay is
approximately equal to the neutrino mass scale
$\< m_\nu \>\approx m^{\mathrm{II}}$
and therefore
neutrinoless double beta decay will be observable in the next round of
experiments if the neutrino mass spectrum is partially degenerate.

\subsection{Theoretical expectations for the Mixing Angles}
In order to discriminate between models, precision measurements of the 
neutrino mixing angles have the potential to play an important role.

One important parameter is the value of the mixing angle
$\theta_{13}$, which is at present only bounded from above to be
smaller than approximately $13^\circ$.  In the type I sequential
dominance case, the mixing angle $\theta_{13}$ is typically of the
order ${\cal O}(m^\mathrm{I}_2/m^{\mathrm{I}}_3)$.  In the
type-II-upgrade scenario this ratio decreases with increasing neutrino
mass scale and is smaller than $\approx 5^\circ$ for partially
degenerate neutrinos even if it was quite large in the
type I limit. Sizable RG corrections, which are usually expected for
partially degenerate neutrinos, are suppressed in the type-II-upgrade
scenario due to small CP phases $\beta_2,\beta_3$ and $\delta$.

Another important parameter is $\theta_{23}$. Its present best-fit value is
close to $45^\circ$, however comparably large deviations are experimentally
allowed as well.   
With sequential dominance, we expect minimal deviations of  
$\theta_{23}$ from $45^\circ$ of the order 
${\cal O}(m^\mathrm{I}_2/m^{\mathrm{I}}_3)$, which could be observed 
by future long-baseline experiments in the type I 
see-saw case.
In the type II upgraded version, the corrections can be significantly smaller
since the ratio $m^\mathrm{I}_2/m^{\mathrm{I}}_3$ decreases with increasing
neutrino mass scale \cite{Antusch:2004xd}. 
For large $\tan \beta$ in the MSSM, the major source for 
the corrections can be RG effects, which are
un-suppressed for small CP phases.

\subsection{Minimal Lepton Flavour Violation} 

At leading order in a mass insertion approximation
the branching fractions of LFV processes are given by
\footnote{The mass insertion approximation given in
  Eq.\ref{eq:BR(li_to_lj)} is for illustrative purposes only.
The conclusions quoted below from \cite{Blazek:2002wq}
do not rely on this approximation.}
\beq
{\rm BR}(l_i \rightarrow l_j \gamma)\approx
        \frac{\alpha^3}{G_F^2}
        f(M_2,\mu,m_{\tilde{\nu}})
        |m_{\tilde{L}_{ij}}^2|^2  \tan ^2 \beta
    \label{eq:BR(li_to_lj)}\; ,
\eeq
where $l_1=e, l_2=\mu , l_3=\tau$,
and where the off-diagonal slepton doublet mass squared is given
in the leading log approximation (LLA) by
\beq
m_{\tilde{L}_{ij}}^{2(LLA)}
\approx -\frac{(3m_0^2+A_0^2)}{8\pi ^2}C_{ij}\; .
\label{lla}
\eeq
With sequential dominance, using the notation of Eqs.\ref{seq1}, 
the leading log coefficients relevant for
$\mu \rightarrow e\gamma$ and $\tau \rightarrow \mu \gamma$
are given approximately as
\bea
C_{21} & = & ab\ln \frac{M_U}{X} +de\ln \frac{M_U}{Y}\; , \nonumber \\
C_{32} & = & bc\ln \frac{M_U}{X} +ef\ln \frac{M_U}{Y}\; .
\label{C2131}
\eea
Large rates of 
$\tau \rightarrow \mu \gamma$ which is the characteristic expectation
of lop-sided models in general \cite{Blazek:2002wq}. 
Such models occur if the dominant right-handed neutrino is the
heaviest one.
A global analysis of LFV has been performed in the constrained
minimal supersymmetric standard model (CMSSM) for the case
of sequential dominance, focussing on the two cases
where the dominant right-handed neutrino is either 
the heaviest one or the lightest one
\cite{Blazek:2002wq}. 
If the dominant right-handed neutrino is the lightest one then
the rate for $\tau \rightarrow \mu \gamma$
is well below observable values. Therefore $\tau \rightarrow \mu
\gamma$ provides a good discriminator between the two cases 
of dominance. In \cite{Blazek:2002wq} it is shown that 
the rate for $\mu \rightarrow e \gamma$ may determine the order
of the sub-dominant neutrino Yukawa couplings in the flavour basis.

\section{Model Building Applications}
\subsection{Effective Two Right-Handed Neutrino Models}
In sequential dominance we have seen that one of the right-handed
neutrinos effectively decouples from the see-saw mechanism.
If the decoupled right-handed neutrino is also the heaviest
then it would be expected to play no part in phenomenology.
In this case sequential
dominance reduces to effectively two right-handed neutrino models
\cite{King:1998jw}.
Recently there have been several studies based on the ``minimal see-saw'' 
involving two right-handed neutrinos \cite{Frampton:2002qc},
and it is worth bearing in mind that such models 
could naturally arise as the limiting case of sequential dominance.

\subsection{Sneutrino Inflation Models}
Sequential dominance has recently also been applied to sneutrino
inflation \cite{Ellis:2003sq}, \cite{Antusch}. 
Requiring a low reheat temperature
after inflation, in order to solve the gravitino problem,
forces the sneutrino inflaton to couple very weakly to 
ordinary matter and its superpartner almost to decouple from the
see-saw mechanism.
This decoupling of a right-handed neutrino from the see-saw mechanism
is a characteristic of sequential dominance.

\subsection{GUT and Family Symmetry Models}
There are many models in the literature based on sequential dominance.
A Pati-Salam model with 
$U(1)$ family symmetry was considered in \cite{King:2000ge}.
Single right-handed neutrino dominance has also been applied to
SO(10) GUT models involving a $U(2)$ family symmetry
\cite{Barbieri:1999pe}. Sequential dominance with
$SU(3)$ family symmetry and $SO(10)$ GUTs has been considered in
\cite{King:2001uz}. Type II up-gradable models
based on sequential dominance of the ISD type 
with $SO(3)$ family symmetry have been considered in 
\cite{Antusch:2004re,Antusch:2004xd}.
For GUT models the renormalisation group corrections need to be taken
into account, although for a natural hierarchy such corrections are
only a few per cent \cite{King:2000hk}.

As an example of a model based on a non-Abelian family symmetry,
we briefly review the model proposed in \cite{King:2003rf}.
The model uses the largest family symmetry $SU(3)$ consistent
with $SO(10)$ GUTs. An important further
motivation for $SU(3)$ family symmetry is, in the framework of
sequential dominance,
to relate the second and third entries of the Yukawa matrix, as required to
obtain an almost maximal 23 mixing in the atmospheric neutrino
sector \cite{King:2001uz}.
The theoretical requirements that
the neutrino Yukawa matrix resembles the quark Yukawa matrices,
and therefore has a large 33 element with no large off-diagonal
elements and a texture zero in the 11 position
leads uniquely to the dominant right-handed neutrino
being the first (lightest) one. Assuming this then
the atmospheric neutrino mixing angle
is given by
$\tan \theta_{23}^{\nu}\approx Y^{\nu}_{21}/Y^{\nu}_{31}\approx 1$.
The sequential dominance conditions which were assumed earlier
will here be derived from the symmetries of the model.
Thus this model provides an example of the application of sequential
dominance to realistic models of flavour, and shows how the conditions
of sequential dominance which were simply assumed earlier can motivate
models based on GUTs and family symmetry which are capable of
explaining these conditions. In other words, the conditions for
sequential dominance can provide clues to the nature of the underlying
flavour theory.

The starting point of the model is the observation that an excellent
fit to all quark data is given by the approximately symmetric form
of quark Yukawa matrices \cite{Roberts:2001zy}
\begin{equation}
Y^{u}\propto \left(
\begin{array}{ccc}
0 & \epsilon ^{3} & O(\epsilon ^{3}) \\
. & \epsilon ^{2} & O(\epsilon ^{2}) \\
. & . & 1%
\end{array}%
\right) ,\ \ \ \ Y^{d}\propto \left(
\begin{array}{crc}
0 & 1.5\bar{\epsilon}^{3} & 0.4\bar{\epsilon}^{3} \\
. & \bar{\epsilon}^{2} & 1.3\bar{\epsilon}^{2} \\
. & . & 1%
\end{array}%
\right)  \label{yuk}
\end{equation}%
where the expansion parameters $\epsilon $ and $\bar{\epsilon}$ are given by
\begin{equation}
\epsilon \approx 0.05,\ \ \bar{\epsilon}\approx 0.15.  \label{exp}
\end{equation}

This motivates a particular model in which
the three families are unified as triplets under
an $SU(3)$ family symmetry, and $16's$ under an $SO(10)$ GUT
\cite{King:2001uz,King:2003rf},
\begin{equation}
\psi_i= (3,16),
\end{equation}
where the $SO(10)$ is broken via the Pati-Salam group resulting in:
\begin{equation}
{\psi_i}=(3,4,2,1),\ \ {\bar{\psi}}_i=(3,\bar{4},1,\bar{2}).
\end{equation}
Further symmetries $R\times Z_2\times U(1)$
are assumed to ensure that
the vacuum alignment leads to a universal form of Dirac mass matrices
for the neutrinos, charged leptons and quarks \cite{King:2003rf}.
To build a viable model we also need spontaneous breaking of the family
symmetry
\begin{equation}
SU(3)\longrightarrow SU(2)\longrightarrow {\rm Nothing}  \label{fsb}
\end{equation}%
To achieve this symmetry breaking additional Higgs fields $\phi
_{3},$ $\overline{\phi }_{3},$ $\phi _{23}$ and $\overline{\phi }_{23}$
are required. The largeness of the third
family fermion masses implies that $SU(3)$ must be strongly broken by new
Higgs antitriplet fields $\phi _{3}$ which develop a vev in the third $SU(3)$
component $<\phi _{3}>^{T}=(0,0,a_{3})$ as in \cite{King:2001uz}.
$\phi _{3}^{i}$ transforms under
$SU(2)_{R}$ as ${\bf 3\oplus 1}$ rather than being $SU(2)_{R}$ singlets as
assumed in \cite{King:2001uz}, and develops vevs in the $SU(3)\times
SU(2)_{R}$ directions
\begin{equation}
<\phi _{3}>=<\overline{\phi _{3}}>=\left(
\begin{array}{c}
0 \\
0 \\
1%
\end{array}%
\right) \otimes \left(
\begin{array}{cc}
a_{3}^{u} & 0 \\
0 & a_{3}^{d}%
\end{array}%
\right) .  \label{phi3vev}
\end{equation}%
The symmetry breaking also involves the $SU(3)$ antitriplets $\phi _{23}$
which develop vevs \cite{King:2001uz}
\begin{equation}
<\phi _{23}>=\left(
\begin{array}{c}
0 \\
1 \\
1%
\end{array}%
\right) b,  \label{phi23vevs}
\end{equation}%
where, as in \cite{King:2001uz}, vacuum alignment ensures that the vevs are
aligned in the 23 direction. Due to D-flatness there must also be
accompanying Higgs triplets such as $\overline{\phi _{23}}$ which develop
vevs \cite{King:2001uz}
\begin{equation}
<\overline{\phi _{23}}>=\left(
\begin{array}{c}
0 \\
1 \\
1%
\end{array}%
\right) b.  \label{phibar23vevs}
\end{equation}%
We also introduce an
adjoint $\Sigma$ field which develops vevs in the $SU(4)_{PS}\times SU(2)_R$
direction which preserves the hypercharge generator $Y=T_{3R}+(B-L)/2$, and
implies that any coupling of the $\Sigma$ to a fermion and a messenger such
as $\Sigma^{a \alpha}_{b \beta}\psi^c_{a\alpha}\chi^{b\beta}$, where the $%
SU(2)_R$ and $SU(4)_{PS}$ indices have been displayed explicitly, is
proportional to the hypercharge $Y$ of the particular fermion component of $%
\psi^c$ times the vev $\sigma$. In addition a $\theta$ field is
required for the construction of Majorana neutrino masses.

The leading operators allowed by the symmetries are
\begin{eqnarray}
P_{{\rm Yuk}} &\sim &\frac{1}{M^{2}}\psi _{i}\phi _{3}^{i}\bar{\psi} _{j}\phi
_{3}^{j}h  \label{op1} \\
&+&\frac{\Sigma }{M^{3}}\psi _{i}\phi _{23}^{i}\bar{\psi} _{j}\phi _{23}^{j}h
\label{op2} \\
P_{{\rm Maj}} &\sim &\frac{1}{M}\bar{\psi} _{i}\theta ^{i}\theta ^{j}
\bar{\psi}_{j}  \label{mop1}
\end{eqnarray}%
where the operator mass scales, generically denoted by $%
M$ may differ and we have suppressed couplings of $O(1).$

The final form of the Yukawa matrices and heavy Majorana matrix
after inserting a particular choice of order unity coefficients is
\cite{King:2003rf}
\begin{eqnarray}
Y^{u} &\approx &\left(
\begin{array}{llr}
0 & 1.2\epsilon ^{3}& 0.9\epsilon ^{3} \\
-1.2\epsilon ^{3} & -\frac{2}{3}\epsilon ^{2} & -\frac{2}{3}\epsilon
^{2} \\
-0.9\epsilon ^{3} & -\frac{2}{3}\epsilon ^{2} & 1
\end{array}%
\right) \bar{\epsilon},  \label{Yu} \\
Y^{d} &\approx &\left(
\begin{array}{llr}
0 & 1.6\bar{\epsilon}^{3} & 0.7\bar{\epsilon}^{3}
\\
-1.6\bar{\epsilon}^{3} & \bar{\epsilon}^{2} & \bar{\epsilon}^{2}+
\bar{\epsilon}^{\frac{5}{2}} \\
-0.7\bar{\epsilon}^{3} & \bar{\epsilon}^{2} & 1
\end{array}%
\right) \bar{\epsilon},  \label{Yd} \\
Y^{e} &\approx &\left(
\begin{array}{llr}
0 & 1.6\bar{\epsilon}^{3} & 0.7\bar{\epsilon}^{3}
\\
-1.6\bar{\epsilon}^{3} & 3\bar{\epsilon}^{2} & 3\bar{\epsilon}^{2}\\
-0.7\bar{\epsilon}^{3} & 3\bar{\epsilon}^{2} & 1
\end{array}%
\right) \bar{\epsilon},  \label{Ye} \\
Y^{\nu } &\approx &\left(
\begin{array}{llr}
0 & 1.2\epsilon ^{3} & 0.9\epsilon ^{3} \\
-1.2\epsilon ^{3} & -\alpha \epsilon ^{2} & -\alpha \epsilon ^{2}\\
-0.9\epsilon ^{3} & -\alpha \epsilon ^{2}-\epsilon ^{3} & 1
\end{array}%
\right) \bar{\epsilon}.  \label{Ynu}
\end{eqnarray}
\begin{equation}
M_{RR}\approx \left(
\begin{array}{ccr}
\epsilon ^{6}\bar{\epsilon}^{3} & 0 & 0 \\
0 & \epsilon ^{6}\bar{\epsilon}^{2} & 0 \\
0 & 0 & 1%
\end{array}%
\right) M_{3}.  \label{MRRA}
\end{equation}

The model gives excellent agreement with the quark and
lepton masses and mixing angles. For the up and down quarks the form of $%
Y^{u}$ and $Y^{d}$ given in Eq.\ref{Yu}, \ref{Yd} is consistent with the
phenomenological fit in Eq.\ref{yuk}.
The charged lepton mass matrix is of the Georgi-Jarslkog \cite{Georgi:1979df}
form which, after
including radiative corrections, gives an excellent description of the
charged lepton masses. In the neutrino sector the parameters satisfy the
conditions of sequential dominance in Eq.\ref{srhnd},
with the lightest right-handed neutrino
giving the dominant contribution to the heaviest physical neutrino mass, and
the second right-handed neutrino giving the leading subdominant
contribution, providing that $\alpha \sim \epsilon $.

Analytic estimates of neutrino masses and mixing angles for sequential
dominance were derived in Section 3
\cite{King:1998jw}.
With the dominant right-handed neutrino of mass $Y$ being the lightest one,
the matrices in Eq.\ref{seq1} should be re-ordered 
as follows before comparing to the neutrino matrices in
Eqs.\ref{Ynu} and \ref{MRRA}:
\begin{equation}
M_\mathrm{RR}=
\left( \begin{array}{ccc}
Y & 0 & 0    \\
0 & X & 0 \\
0 & 0 & Z
\end{array}
\right), \ \ \ \ 
m^\nu_\mathrm{LR}=Y^{\nu }v_2=
\left( \begin{array}{ccc}
0 & a & p    \\
e & b & q\\
f & c & r
\end{array}
\right).
\label{seq2}
\end{equation}
The above re-ordering of course leaves the results in Eqs.\ref{masses} and
\ref{angles} unchanged.
The analytic estimates for the neutrino masses are obtained
by comparing the matrices in Eqs.\ref{seq2} to those in
Eqs.\ref{Ynu} and \ref{MRRA} then using 
the results in Eqs.\ref{masses} and \ref{angles}:
\begin{eqnarray}
m_{1} &\sim &\bar{\epsilon}^{2}\frac{v_{2}^{2}}{M_{3}} \\
m_{2} &\approx & 5.8\frac{v_{2}^{2}}{M_{3}} \\
m_{3} &\approx &15\frac{v_{2}^{2}}{M_{3}}
\end{eqnarray}%
and neutrino mixing angles:
\begin{eqnarray}
\tan \theta _{23}^{\nu } &\approx &  1.3 \\
\tan \theta _{12}^{\nu } &\approx &  0.66 \\
\theta _{13}^{\nu } &\approx &1.6\bar{\epsilon}
\end{eqnarray}%
The physical lepton mixing angle $\theta_{13}$ receives
a large contribution from the neutrino sector $\theta_{13}^{\nu}\sim 0.3$
at the high energy scale, for this choice of parameters,
compared to the current CHOOZ limit
$\theta_{13}\leq 0.2$. However the physical
mixing angles will receive charged lepton contributions
\cite{King:1998jw} and all
the parameters are subject to radiative corrections in running from the high
energy scale to low energies, although in sequential dominance
models these corrections are only a few per cent
\cite{King:2000hk}. Thus the
model predicts that $\theta_{13}$ is close
to the current CHOOZ limit, and could be observed by
the next generation of long baseline experiments such as MINOS or OPERA.

\section{Leptogenesis}
Neutrino mass allows the possibility that the baryon asymmetry of the
universe is generated by out-of-equilibrium decay of lepton-number
violating Majorana right-handed
(s)neutrinos, whose decays result in a net lepton number which is subsequently
converted to a net baryon number by sphaleron transitions.
This mechanism is known as leptogenesis \cite{yanagida1}.
In models which give a natural neutrino
mass hierarchy, if the dominant right-handed neutrino is the lightest
one then the washout parameter $\tilde{m_1}\sim O(m_3)$, 
which is rather too large compared to the optimal value
of around $10^{-3}$ eV,
while if the dominant right-handed neutrino is either the
intermediate one or the heaviest one then one finds
$\tilde{m_1}\sim O(m_2)$ or arbitrary $\tilde{m_1}$, which can be 
closer to the desired value \cite{Hirsch:2001dg}. 

It is interesting to note that if the dominant right-handed
neutrino is the lightest one, and there is a 11 texture zero,
as is the case in the $SU(3)\times SO(10)$ model discussed in the
last section, then 
there is a link between the CP violation required for leptogenesis,
$\phi_{\rm COSMO}$,
and the phase $\delta$ measurable in accurate neutrino oscillation
experiments \cite{King:2002qh}.
$\delta$ turns out to be a function of
the same two see-saw phases that also determine $\phi_{\rm COSMO}$.
If both the see-saw phases are zero,
then both $\phi_{\rm COSMO}$ and $\delta$ are necessarily zero.
This feature is absolutely crucial. It means that,
barring cancellations, measurement of a non-zero value for
the phase $\delta$ at a neutrino factory will be a signal of a
non-zero value of the leptogenesis phase $\phi_{\rm COSMO}$.
We also find the remarkable result
\beq
|\phi_{\rm COSMO}|=|\phi_{\beta \beta 0\nu}|.
\label{remarkable2}
\eeq
where $\phi_{\beta \beta 0\nu}$ is the phase which enters the rate
for neutrinoless double beta decay \cite{King:2002qh}.

\begin{figure}
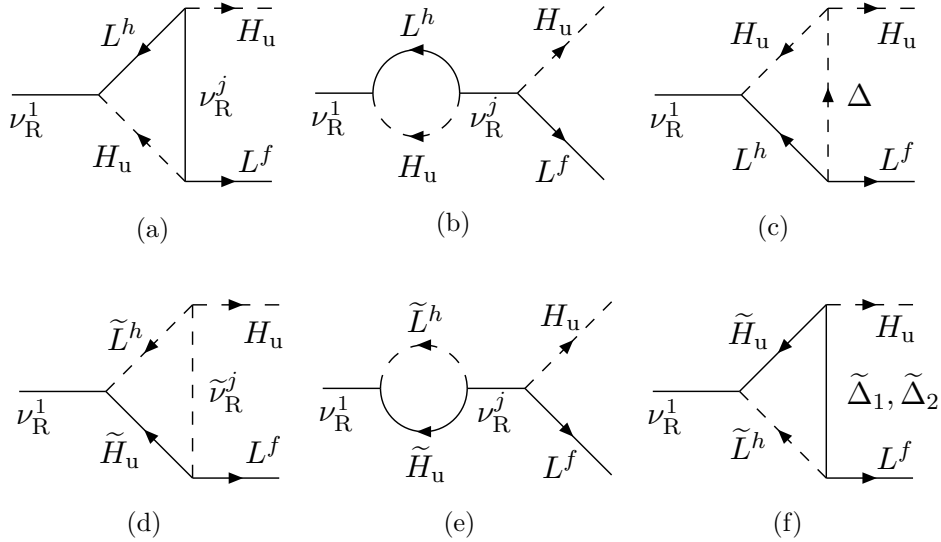

 \begin{center}  
 \CenterEps[1]{TypeIILoops}
 \end{center}
 \caption{\label{fig:Leptogenesis_Triplet_SUSY} Loop diagrams in the
 MSSM which contribute to the decay $\nu^1_{\mathrm{R}}\rightarrow
 L^f_a H_\mathrm{u}{}_b$ for the case of a type II see-saw mechanism
 where the direct mass term for the neutrinos stems from the induced
 vev of a Higgs triplet.  In diagram (f), $\widetilde \Delta_1$ and
 $\widetilde \Delta_2$ are the mass eigenstates corresponding to the
 superpartners of the SU(2)$_{\mathrm{L}}$-triplet scalar fields
 $\Delta$ and $\bar{\Delta}$.  The SM diagrams are the ones where no
 superpartners (marked by a tilde) are involved and where
 $H_\mathrm{u}{}$ is renamed to the SM Higgs.  }
\end{figure}

We now discuss the consequences of the neutrino mass scale for
leptogenesis via the out-of-equilibrium decay of the lightest
right-handed (s)neutrinos in type II see-saw models
\cite{Antusch:2004xy}.  
In \cite{Antusch:2004xy} we 
calculated the type II contributions to the decay asymmetries for
minimal scenarios based on the Standard Model (SM) and on the Minimal
Supersymmetric Standard Model (MSSM), where the additional direct mass
term for the neutrinos stems from the induced vev of a triplet Higgs.
The diagrams are shown in Figure \ref{fig:Leptogenesis_Triplet_SUSY}.
The result we obtained 
for the supersymmetric case is new and we corrected the
previous result in the scenario based on the Standard Model.  
\cite{Hambye:2003ka}.

We subsequently derived a general upper bound on the decay asymmetry and
found that it increases with the neutrino mass scale:
\begin{subequations}\label{eq:BoundsEffDecayAss}\begin{eqnarray}\label{eq:BoundEffDecayAss_SM}
|\varepsilon^{\mathrm{SM}}_{1}| 
&\le&
  \frac{3}{16 \pi} 
 \frac{M_{\mathrm{R}1}}{v_\mathrm{u}^2}  m^\nu_\mathrm{max}\; ,
 \\
\label{eq:BoundEffDecayAss_MSSM} |\varepsilon^{\mathrm{MSSM}}_{1}| 
&\le&
  \frac{3}{8 \pi} 
 \frac{M_{\mathrm{R}1}}{v_\mathrm{u}^2}  m^\nu_\mathrm{max} 
\end{eqnarray}\end{subequations}

It leads to a
lower bound on the mass of the lightest right-handed neutrino, which
is significantly below the type I bound for partially degenerate
neutrinos.  It is worth emphasizing that these results are in sharp
contrast to the type I see-saw mechanism where an upper bound on the
neutrino mass scale is predicted.  Here we find no upper limit on the
neutrino mass scale which may be increased arbitrarily. Indeed we find
that the lower bound on the mass of the lightest right-handed neutrino
decreases as the physical neutrino mass scale increases.  This allows
a lower reheat temperature, making thermal leptogenesis more
consistent with the gravitino constraints in supersymmetric models.

\section{Conclusion}
In this talk
we have discussed how a natural neutrino mass hierarchy can follow from
the type I see-saw mechanism, and a natural neutrino mass 
degeneracy from the type II see-saw mechanism, where the bi-large
mixing angles can arise from either the neutrino or charged lepton
sector. The key to achieving naturalness is the idea of sequential
dominance of right-handed neutrinos, namely that in the see-saw
mechanism one of the right-handed neutrinos dominates and
couples with approximately 
equal strength to the $\tau$ and $\mu$ families, leading
to an approximately maximal atmospheric mixing angle.
A second right-handed neutrino then plays the leading 
sub-dominant role, and couples with approximately equal strength to
all three families, leading to a large solar mixing angle.
We have shown that this can lead either to a hierchical
neutrino mass spectrum or, if a type II contribution proportional
to the unit matrix is considered, to approximately
degenerate neutrino masses. We have summarised the 
phenomenological implications of such a natural approach to 
neutrino model building, and have discussed some of the 
model building applications, focussing on the $SU(3)\times SO(10)$
model. We then turned to leptogenesis and mentioned the possible
link between the leptogenesis phase and the phase measurable in
neutrino oscillation experiments. We then pointed out that
in natural type II models the leptogenesis asymmetry parameter becomes
proportional to the neutrino mass scale, in sharp contrast to the
type I case, which leads to an upper bound on the neutrino mass scale,
allowing lighter right-handed neutrinos and hence making leptogenesis
more consistent with the gravitino constraints in supersymmetric models.

\bigskip\noindent
{\large \bf Acknowledgements}
I would like to thank Stefan Antusch for his collaboration in some 
of the work presented here. I would also like to thank the organisers
of Pascos '04 and the Nobel Symposium 129.

\end{document}